\documentclass[a4paper]{jpconf}
\usepackage{amsmath}
\usepackage{graphicx}
\usepackage{natbib}
\usepackage{hyperref}
\usepackage{cite}
\usepackage{bm}
\usepackage{rotating}
\usepackage{tabularx}
\usepackage{ulem}
\usepackage{float}
\usepackage{multirow}
\usepackage{siunitx}
\usepackage{booktabs}
\usepackage{enumerate}
\begin{document}
\title{Scaling with deformation in probable $p$-wave halo $^{34}$Na}

\author{Manju$^{1a}$, Jagjit Singh$^2$, Shubhchintak$^{1,3}$, and R. Chatterjee$^{1b}$}

\address{$^1$Department of Physics, Indian Institute of Technology Roorkee, Uttarakhand- 247667, India}

\address{$^2$Research Centre for Nuclear Physics (RCNP), Osaka University, Ibaraki 567-0047, Japan}

\address{$^3$Physique Nucl\'{e}aire Th\'{e}orique et Physique Math\'{e}matique, C. P. 229, Universit\'{e} Libre de Bruxelles (ULB), B 1050 Brussels, Belgium}
\ead{manju@ph.iitr.ac.in$^a$, rchatterjee@ph.iitr.ac.in$^b$}

\begin{abstract}
We investigate the electric dipole response of $^{34}$Na, a probable $p$-wave one-neutron halo nucleus, lying in the ``island of inversion" and having a deformed structure. We use a semi-analytic approach to probe the dipole response and further compare the results obtained from a post form finite-range distorted wave Born approximation theory of Coulomb breakup. The effects of deformation are figured out on the peak positions of the electric dipole strength distribution which further constraint the one-neutron separation energy of the deformed projectile and it leads to a two-dimensional scaling of total $B(E1)$ strength with parameters: separation energy and deformation. 
\end{abstract}

\section{Introduction}
The advancements in radioactive ion beams (RIB) have created a new domain of research in nuclear physics by exploring the possibility of nuclei far from the limits of stability. Experiments with these RIBs lead to the discovery of exotic nuclei and in particular the new phenomena, halo, in which one, or two, valence nucleons decouple from a tightly bound core resulting in a cluster like structure \citep{Tanihata1985}. The conventional nuclear structure methods can not be used to study these nuclei due to their short half lives and small one or two nucleon separation energies. Breakup of these short lived and weakly bound projectile on heavy targets could be a convenient tool to investigate their structure. Coulomb breakup of halo nuclei is of particular interest due to their large breakup cross-sections which is a result of the dominance of electric dipole responses at low excitation energies, called soft $E1$ excitation. The electric dipole strength for lighter halo nuclei, such as $^{6}$He, $^{11}$Li, $^{11}$Be and $^{19}$C has been studied prodigiously both theoretically as well experimentally \citep{Aumann1999, Sackett1993, Nakamura1994, Nakamura1999}. Nevertheless, the experimental data for $E1$ strength does not exist for nuclei lying in the medium mass region and therefore we extend this research to deformed nuclei lying in medium mass region of ``island of inversion" around $N=20$.

In this direction, recently, we have studied the electric dipole responses of medium mass $p$-wave halos $^{31}$Ne, $^{34}$Na and $^{37}$Mg, using two different theoretical approaches \citep{Manju2019}. One is based on the asymptotic counterparts where bound and continuum wave functions are given in terms of asymptotic forms of spherical Hankel and Bessel functions, respectively, called analytical model \citep{Nagarajan2005} and the other is the post form theory of Coulomb dissociation (CD) under the aegis of a finite-range distorted wave Born approximation (FRDWBA) \citep{Chatterjee2018}. Apart from this, we estimated the one-neutron separation energy of $^{31}$Ne, $^{34}$Na and $^{37}$Mg, and the results obtained are in good agreement with the available theoretical estimates and experimental data. These results are based on the assumption of spherical projectiles. As an extension of this study, we investigate the behavior of electric dipole response functions of $p$-wave halos at finite deformations. 
The present study is devoted to $p$-wave halo candidate $^{34}$Na, having deformed structure and lying in the region around $N = 20$ \citep{Singh16}.
A semi-analytic approach (SA), which incorporates the effects of deformation in the bound state wave function, is used to calculate the electric dipole strength and corresponding results are further compared with those obtained from the FRDWBA theory of the CD. FRDWBA has target-projectile electromagnetic interaction at all orders and includes contributions from the entire non-resonant continuum. However, a first-order SA approach resolves the problem of fixing the continuum state which is not an easy task for any nuclear potential model.

We organize the paper as follows:
Section~\ref{fr} briefly describes the formalism where we discuss the FRDWBA theory of breakup reactions and its extension to include the deformation in the ground state of the projectile and Section~\ref{sm} shows the SA approach for deformed bound wave function. In Section~\ref{rnd} the results of dipole response for medium mass halo $^{34}$Na is presented with detailed discussion about effects of projectile deformation. 
Then, in Section~\ref{con} the conclusions are given. 

\section{Mathematical Formulation}\label{mf}

\subsection{Finite-range distorted wave Born approximation theory of Coulomb dissociation}
\label{fr}

Consider a beam of projectile `$a$' impinging on a heavy target `$t$' at considerably high energy breaks up into the two constituents, core `$d$' and valence neutron `$n$'. In the present case $a$, $d$ and $t$ correspond to $^{34}$Na, $^{33}$Na and $^{208}$Pb, respectively. The triple differential cross-section for this reaction is given by,
\begin{eqnarray}
\dfrac{d^3\sigma}{dE_{d} d\Omega_d d\Omega_n} = \dfrac{2 \pi}{\hbar \upsilon_{at}} \rho \sum_{\ell m} \vert \beta_{\ell m} \vert^2,
\label{eqn1}
\end{eqnarray} 
where $\upsilon_{at}$ is the relative velocity of $a-t$ system in the initial channel and $\rho$ is the appropriate three-body phase space factor \citep{Fuchs1982}. The $\beta_{\ell m}$ is the reduced transition amplitude under the post form FRDWBA theory of CD and can be given as,
\begin{eqnarray}
\hat{\ell}\beta_{\ell m}=\int d\textbf{r}_{i} e^{-i \delta \textbf{q}_n.\textbf{r}_i} \chi_{d}^{(-)*}(\textbf{q}_d, \textbf{r}_\textbf{i}) \chi_{a}^{(+)}(\textbf{q}_a, \textbf{r}_i) \int d\textbf{r}_1 e^{-i(\gamma \textbf{q}_n-\alpha\textbf{K}).\textbf{r}_1} V_{dn}(\textbf{r}_1)\phi_a^{\ell m}(\textbf{r}_1).
\label{eqn2}
\end{eqnarray}
\textbf{K} is an effective local momentum of core-target relative system and \textbf{q}$_i$ ($i = a, d, n$) being the Jacobi coordinates of the respective particles. $\alpha$, $\gamma$, and $\delta$ are the mass factors related to the three-body Jacobi coordinate system (Fig.1 in Ref.\citep{Shubh2014}). $\chi$'s represents the distorted waves with incoming (-) and outgoing (+) wave boundary conditions.  
The $\phi_a^{\ell m}(\textbf{r}_1)$ in Eq.(\ref{eqn2}) is the ground state wave function of the projectile which is the only input to the theory. $V_{dn}(\textbf{r}_1)$ is the interaction between the fragments $`d$' and $`n$' in the initial channel which is axially symmetric quadrupole-deformed Woods-Saxon type potential. More detail about FRDWBA formalism can be found in Ref.\citep{Chatterjee2018, Shubh2014, Shubh15}.

Eq.(\ref{eqn1}) computes the relative energy spectra on multiplication with a suitable Jacobian \citep{Fuchs1982}. Further, this relative energy spectra can be represented in terms of dipole response function as \citep{Nakamura1999}
\begin{eqnarray}
\dfrac{d\sigma}{dE_{c}} = \dfrac{16 \pi^3}{9 \hbar c} n_{E1}\dfrac{dB(E1)}{dE_{c}}.
\label{eqn4}
\end{eqnarray}
where $n_{E1}$ is the virtual photon number for the $E1$ excitation. We will compare these FRDWBA results with another approach called a semi-analytical method. 
 
\subsection{Semi-analytical approach}
\label{sm}

In the direct breakup mechanism, the electric dipole strength distribution is the transition matrix element from the bound state with separation energy $S_{n}$ to a continuum state with energy $E_{c}$ \citep{Nakamura1994, Bertulani1992, Bertulanii1992},
\begin{eqnarray}
\dfrac{dB(E1)}{dE_{c}} = (3/4\pi)(Z_{\rm eff}^{(1)}e)^{2} \langle\ell010|\ell^{'}0\rangle^{2} \Bigg\vert\int \mathrm{d}r\ \phi_{b}(r)\ \phi_{c}(E_{c},r)r^{3}\Bigg\vert^{2}.
\label{eqn5}
\end{eqnarray}
The $\phi_b(r)$ in Eq.~(\ref{eqn5}) is the bound state wave function which can be calculated by solving the Schr\"odingler equation containing axially symmetric quadrupole-deformed Woods-Saxon potential. The effective charge, $Z_{\rm eff}^{(\lambda)} e$ for a given multipolarity ${\lambda}$ is defined in Ref.\citep{Typel2005}. We replace the continuum wave function in Eq.~(\ref{eqn5})  by spherical Bessel function which is an exact asymptotic form, details about which can be found in Refs. \citep{Manju2019, Nagarajan2005}. In the breakup processes, most of the contribution in the integral of  Eq.~(\ref{eqn5}) comes from the asymptotic region. 

The axially-symmetric quadrupole-deformed  Woods-Saxon potential along with the spin-orbit term is given by,
\begin{eqnarray}
V_{dn}&=&\Bigg[V^{0}_{ws}\Bigg(1-\beta_2RY_{2}^{0}(\hat{\textbf{r}}_1)\frac{d}{dr_1}\Bigg)+V_s \vec{\ell}.\vec{s} \frac{1}{r_1} \frac{d}{dr_1}\Bigg]f(r_1),\nonumber\\
\label{vdn}
\end{eqnarray} 
where $f(r_1)=\Bigg[1+exp\Bigg(\dfrac{r_1-R}{a_0}\Bigg)\Bigg]^{-1}$ and $R=r_0 A^{1/3}$. $V_{ws}^{0}$ is the depth of the Woods-Saxon potential and $V_s$ is the spin-orbit strength. For more detail on potential parameters see next section.
The dipole response function $B(E1)$ (Eq.~(\ref{eqn5})) can be further used to study the various scaling relations in the breakup of medium mass $p$-wave halos.

\section{Results and discussion}
\label{rnd}
The electric dipole response of medium mass $p$-wave deformed halo $^{34}$Na is investigated by employing the formulation described in Section~\ref{mf}. A semi-analytic approach is used to calculate the total integrated dipole strength ($E1$) and their comparison has been made with FRDWBA theory of Coulomb dissociation at finite deformations. 
\begin{figure}[h!]
	\centering
	\includegraphics[scale=0.3]{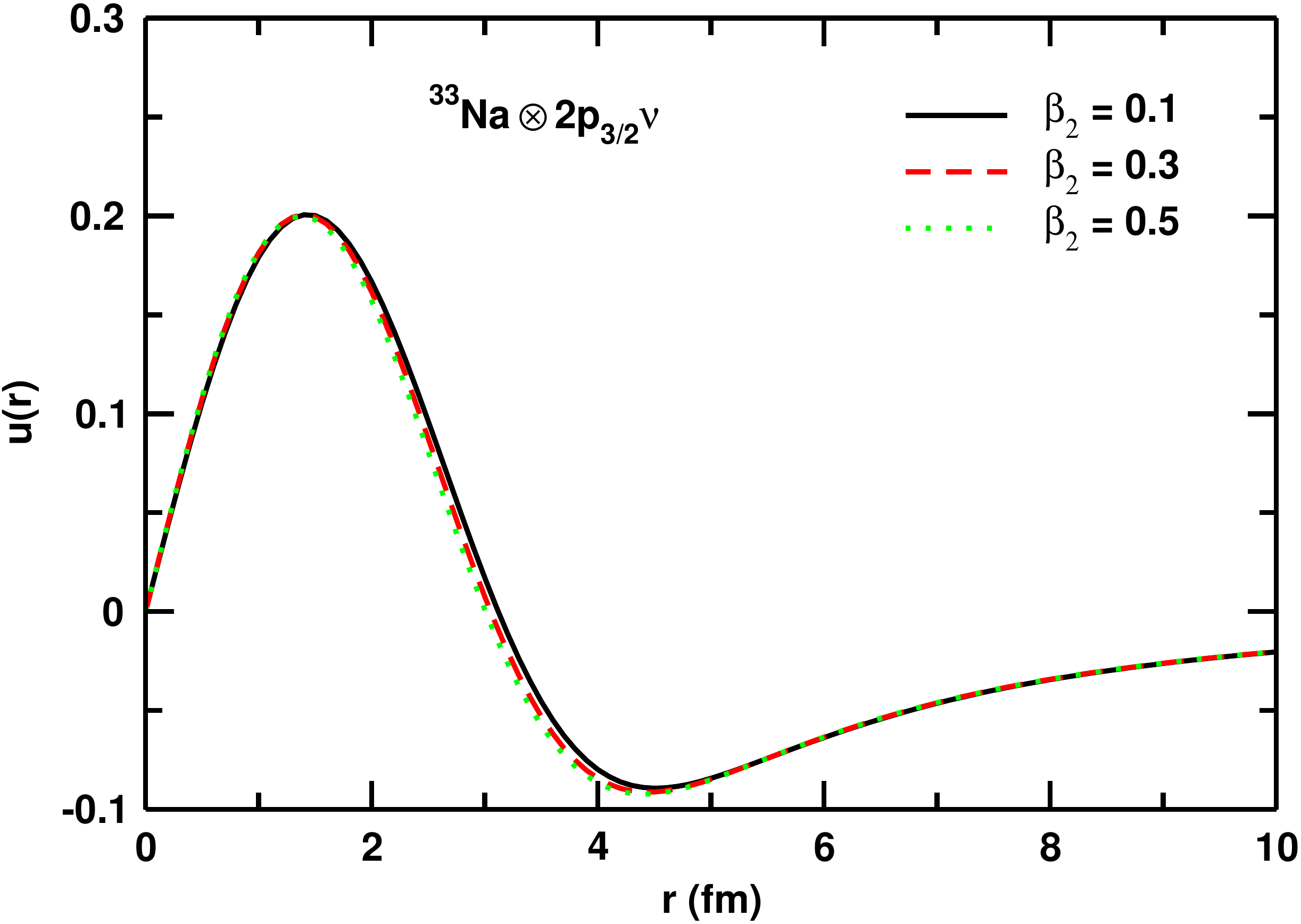}
	\caption{Comparison of the wave function calculated by solving the coupled channel Schr\"odinger equation with Woods-Saxon potential. The $S_n$ is taken to be $0.17$ MeV. These wave functions for a given $\beta_2$ include all components corresponding to $\ell=1,3$ with all the allowed $j$ values. These bound state wave functions are the inputs to our SA model calculations.}
	\label{fig1}
\end{figure}

\begin{figure}[h!]
	\centering
	\includegraphics[scale=0.3]{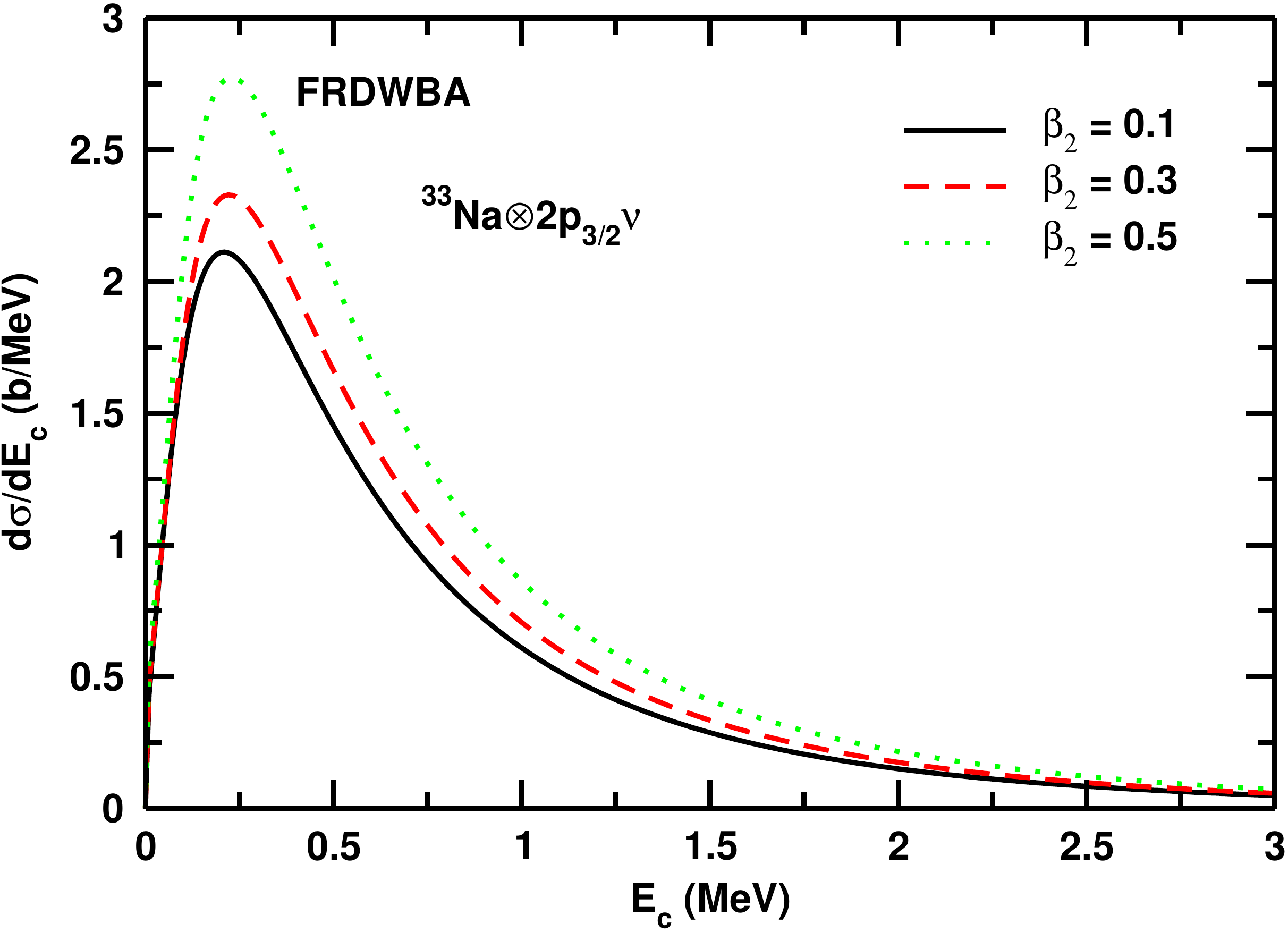}
	\caption{The relative energy spectra in the elastic Coulomb breakup of $^{34}$Na on $^{208}$Pb at $100$\,MeV/u beam energy calculated for three different values of $\beta_2$.}
	\label{fig2}
\end{figure}

As discussed in Refs.\citep{Singh16, Manju2019}, the valance neutron in the ground state of $^{34}$Na is considered in $2p_{3/2}$ orbital. Nevertheless, in order to take the deformation into account we solve the coupled differential Sch\"ordinger equation with the deformed Woods-Saxon potential [Eq.~(\ref{vdn})] to get the wave function. With radius ($r_0$)  and diffuseness ($a_0$) parameters of $1.24$\,fm and $0.62$\,fm, respectively, the depth of the potential is adjusted to reproduce the $S_n$ ($0.17$\.MeV \citep{Gaud12}) of the ground state of $^{34}$Na. The spin-orbit strength ($V_s$) in our calculations is fixed to $15$\,MeV. In Fig.~\ref{fig1}, we present the wave functions for $\beta_2=0.1$ (solid line), $0.3$ (dashed line), and $0.5$ (dotted line). At every deformation, we have included all the components corresponding to $\ell=1,3$ with all the allowed $j$ values. All these wave functions are normalized to unity. We observe that the solid, dotted and dashed curves are almost identical and also, the difference between these curves is negligible. This indicates that the deformation effects on the wave functions are minimal. These deformed wave functions will act as an input to a semi-analytic approach.

The radial bound state wave function used in  Eq.~(\ref{eqn2}) has an admixture of other $\ell$ components of the same parity, for a given $\ell$. However, it was verified in Ref. \citep{Hamamoto2004}, if the binding energies are not too large, the lowest $\ell$ component dominates in the neutron orbits of the deformed potential, irrespective of the size of the deformation. Therefore, we have approximated the radial wave function in Eq.~(\ref{eqn2}), as a single $\ell$ component ($2p_{3/2}$, in this case), which is the solution of Schr\"odinger equation with spherical Woods-Saxon potential. This approximation results in simplification of our FRDWBA calculations. Nevertheless, it is important to mention here that the effects of deformation in the FRDWBA are entering via the potential $V_{dn}$.
In Fig.~\ref{fig2}, we show the relative energy spectra of elastic Coulomb breakup of $^{34}$Na on $^{208}$Pb target at $100$\,MeV/u into fragments $^{33}$Na and valence neutron $n$ at $S_n=0.17$\,MeV. The peak position of the relative energy is observed to be sensitive to the quadrupole-deformation parameter $\beta_2$ but the shape of the curve remains the same independent of the deformation.

\begin{figure}[h!]
	\centering
	\includegraphics[scale=0.35]{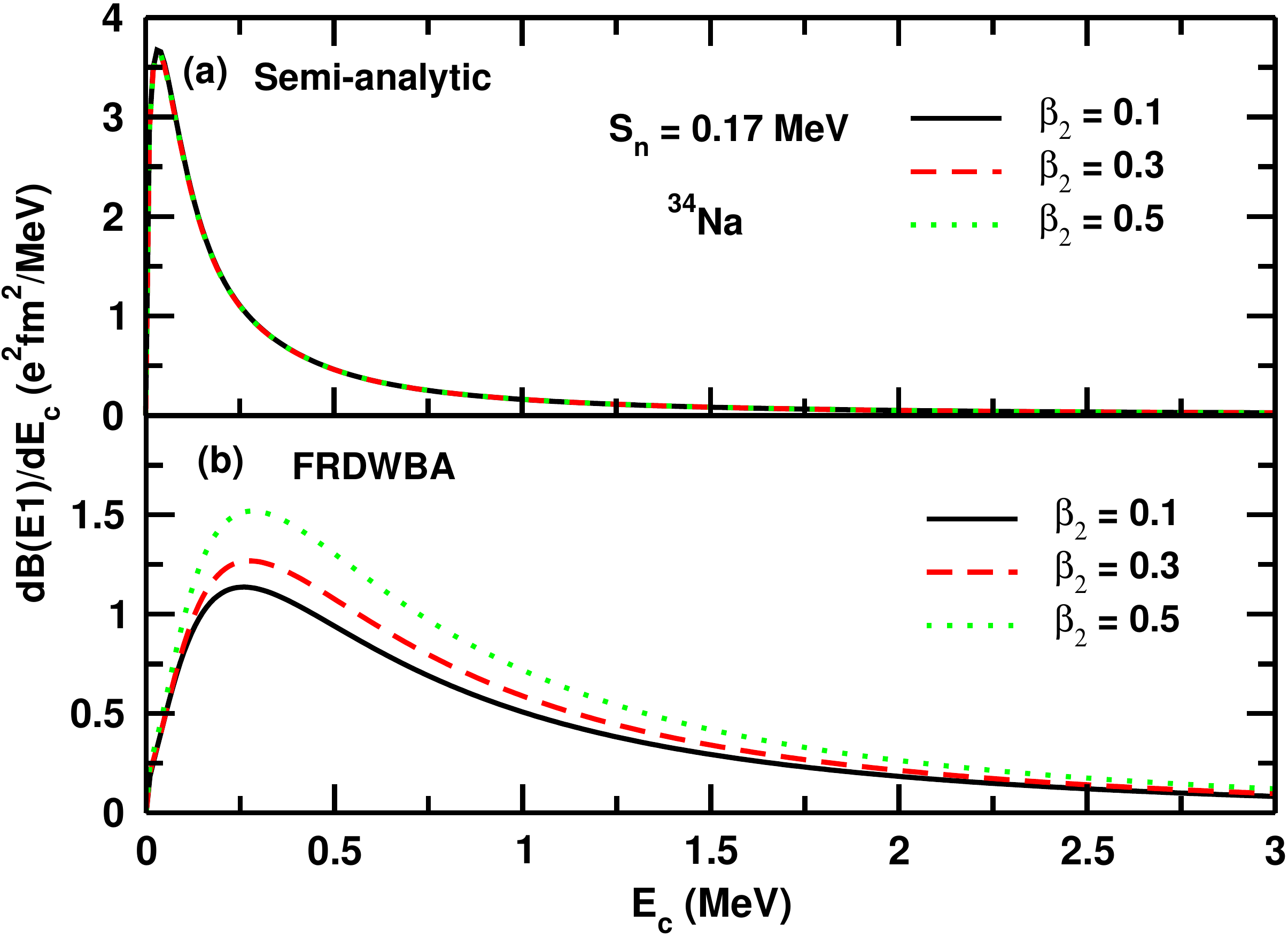}
	\caption{Dipole strength distribution of $^{34}$Na at various deformations (a) within SA approach and (b) within FRDWBA theory of CD.}
	\label{fig3}
\end{figure}

We then calculate the electric dipole response of $^{34}$Na using both the models discussed above. In Fig.~\ref{fig3}, we plot the dipole strength distribution calculated within (a) SA approach and (b) within FRDWBA theory for three different values of $\beta_2$. In the FRDWBA theory we extract dominant dipole strength distribution from the relative energy spectra containing full non-resonant continuum, whereas in the SA approach we sum up the contributions from the predominant deformed $p$-wave bound state to both $s$- and $d$-state continuum. As averred above, the deformed wave function in the SA model do not change much with the deformation (clearly visible in Fig.~\ref{fig1}) especially in the asymptotic region, and hence, one can notice in Fig.~\ref{fig3}(a), the dipole response is weakly dependent on the projectile deformation and distributions appear identical at each deformation. Contrary to this, the dipole response obtained in the FRDWBA theory shows its dependence on $\beta_2$, especially near the peak positions of the response curve. The total integrated $B(E1)$ values obtained corresponding to both the theories are compiled in Table.\ref{be} for $^{34}$Na.  

The comparison of the dipole response function for $S_n=0.17$\,MeV at different $\beta_2$ for both the FRDWBA and semi-analytic approaches suggests that the breakup is entirely $E1$ dominated and higher multipoles other than dipole contribute a little in the breakup process. The total integrated $B(E1)$ values computed from the SA approach are close to those predicted from the FRDWBA theory, especially at small $\beta_2$.

\begin{table}[h]
	\caption{\label{jfonts} Comparison of total integrated $B(E1)$ values of $^{34}$Na from the FRDWBA and the SA approaches.}
	\begin{center}
	\begin{tabular}{lSS}
		\hline\hline
		\toprule
		\multirow{2}{*}{$\beta_2$}~~~~~~~~~~~~~ &
		\multicolumn{2}{c}{$B(E1)$ ($e^2fm^2$)}~~~~~~~~ \\
		& {Semi-analytic}~~~~~ & {FRDWBA} \\
		\midrule
		0.1~~~ & ~~~~~1.00~~~ & ~~~1.25  \\
		0.3~~~ & ~~~~~1.00~~~ & ~~~1.42  \\
		0.5~~~ & ~~~~~1.00~~~ & ~~~1.73  \\
		\bottomrule
		\hline\hline
	\end{tabular}
\end{center}
	\label{be} 
\end{table}

\section{Conclusion}
\label{con}
In summary, we have used a first-order semi-analytical method to compute the total integrated dipole strength $B(E1)$ involving weakly bound medium mass $p$-wave halo nuclei. In semi-classical perturbative models constructing the continuum states is always an issue (i.e. one needs to put addition criterion other than just fixing the ground state). The first order analytical model overcomes this problem by replacing the bound and continuum wave functions with their asymptotic forms. Very recently, with correction factor to the analytical model, $B(E1)$ values have been calculated and estimation of one-neutron separation energies have been made \citep{Manju2019}. Here we extend this work for finite deformation within the semi-analytic model. A more realistic and conventional approach, FRDWBA is used to compare with the semi-analytic estimates taking into account the deformation effects. In the present case, the SA approach contains the sum of contributions of deformed bound $p$-state to continuum $s$- and $d$-states. Furthermore, the deformation is entering in both the theories in a separate manner. In the FRDWBA theory, it is incorporated via a deformed potential whereas, in the SA model deformation enters through the bound state wave function of the projectile. 

We emphasize that dipole strength distribution results are weakly dependent on the deformation in SA approach. We have analyzed that the total $B(E1)$ strength evaluated from SA approach and notice a good agreement with FRDWBA data for $p$-wave halo $^{34}$Na. The results obtained from both the theories suggests that breakup is essentially $E1$ dominated and no higher multipoles dominate other than the dipole. 

In order to understand the underlying physics or reason for difference in shapes of dipole response and total integrated $B(E1)$ values, needs further investigation. We will report this issue along with other $p$-wave halo systems $^{31}$Ne and $^{37}$Mg and some observables such as, ANC's (asymptotic normalization coefficients), root mean square radius, sum rules, \textit{etc}, at finite $\beta_2$ soon elsewhere.

\section{Acknowledgments}
Manju acknowledges financial support from Ministry of Education (Govt. of India) for a doctoral fellowship. This text also presents results from research supported by the Scheme for Promotion of Academic and Research
Collaboration (SPARC/2018-2019/P309/SL), MHRD, Govt. of India.


\begin{thebibliography}{20}
\bibitem[Tanihata et al. (1985)]{Tanihata1985}
Tanihata I, \textit{et al.} 1985 \textit{Phys. Lett.} B \textbf{160} 380.

\bibitem[Aumann et al. (1999)]{Aumann1999}
Aumann T, \textit{et al.} 1999 \textit{Phys. Rev.} C \textbf{59} 1252.

\bibitem[Sackett et al. (1993)]{Sackett1993}
Sackett D, \textit{et al.} 1993 \textit{Phys. Rev}. C \textbf{48} 118; Shimoura S, \textit{et al.} 1995 \textit{Phys. Lett.} B \textbf{348} 29; Zinser M, \textbf{et al.}1997 \textit{Nucl. Phys.} A \textbf{619} 151.

\bibitem[Nakamura et al. (1994)]{Nakamura1994}
Nakamura T, \textit{et al.} 1994 \textit{Phys. Lett}. B \textbf{331} 296.

\bibitem[Nakamura et al. (1999)]{Nakamura1999}
Nakamura T, \textit{et al.} 1999 \textit{Phys. Rev. Lett.} \textbf{83} 1112.

\bibitem[Manju et al. (2019)]{Manju2019}
Manju, Singh J, Shubhchintak and Chatterjee R 2019 \textit{Euro. Phys. J} A \textbf{55} 5.

\bibitem[Nagarajan et al. (2005)]{Nagarajan2005}
Nagarajan M A, Lenzi S M, and Vitturi A 2005 \textit{Eur. Phys. J.} A \textbf{24} 63.

\bibitem[Chatterjee et al. (2018)]{Chatterjee2018}
Chatterjee R and Shyam R 2018 \textit{Prog. Part. Nucl. Phys.} \textbf{103} 67.

\bibitem[Singh et al. (2016)]{Singh16}
Singh G, Shubhchintak, and Chatterjee R 2016 \textit{Phys. Rev. C} \textbf{94} 024606.


\bibitem[Fuchs et al. (1982)]{Fuchs1982}
Fuchs H 1982 \textit{Nucl. Instrum. Methods} \textbf{200} 361.
\bibitem[Shubh et al. (2014)]{Shubh2014}
Shubhchintak, and Chatterjee R 2014 \textit{Nucl. Phys.} A \textbf{922} 99.

\bibitem[Shubh et al. (2015)]{Shubh15}
Shubhchintak, Neelam, Chatterjee R, Shyam R, and Tsushima K 29015 \textit{Nucl. Phys.} A \textbf{939} 101.

\bibitem[Bertulani et al. (1992)]{Bertulani1992}
Bertulani C A and Canto L F 1992 \textit{Nucl. Phys.} A \textbf{539} 163.
\bibitem[Bertulani et al. (1992)]{Bertulanii1992}
Bertulani C A and Sustich A 1992 \textit{Phys. Rev.} C \textbf{46} 6.
\bibitem[Typel et al. (2005)]{Typel2005}
Typel S and Baur G 2005 \textit{Nucl. Phys.} A \textbf{759} 247.

\bibitem[Gaud et al. (2012)]{Gaud12}
Gaudefroy L, \textit{et al.,} 2012\textit{ Phys. Rev. Lett}.\textbf{109} 202503.

\bibitem[Hamamoto et al. (2004)]{Hamamoto2004}
Hamamoto I 2004 \textit{Phys. Rev}. C \textbf{69} 041306(R).


\end{thebibliography}
\end{document}